\documentclass[aps,pre,twocolumn,showpacs,superscriptaddress,groupedaddress]{revtex4-1}
\usepackage{graphicx}  
\usepackage{dcolumn}   
\usepackage{bm}        	
\usepackage{amssymb}   
\usepackage{amsmath}

\usepackage{natbib}

\begin{document}

\title{Efficiency of Monte Carlo Sampling in Chaotic Systems}

\author{Jorge C. Leit\~ao}
\email[Author to whom correspondence should be sent. E-mail address:
      ]{jleitao@pks.mpg.de}
\affiliation{Max Planck Institute for the Physics of Complex Systems, 01187 Dresden, Germany}

\author{J. M. Viana Parente Lopes}
\affiliation{Department of Physics and Center of Physics, University of Minho, P-4710-057, Braga, Portugal}
\affiliation{Physics Engineering Department, Engineering Faculty of the University of Porto, 4200-465 Porto, Portugal}

\author{Eduardo G. Altmann}
\affiliation{Max Planck Institute for the Physics of Complex Systems, 01187 Dresden, Germany}

\date{\today}

\begin{abstract}
In this paper we investigate how the complexity of chaotic phase spaces affect the efficiency of importance sampling Monte Carlo simulations.
We focus on a flat-histogram simulation of the distribution of finite-time Lyapunov exponent in a simple chaotic system and obtain analytically that the computational effort of the simulation:
(i) scales polynomially with the finite-time, a tremendous improvement over the exponential scaling obtained in usual uniform sampling simulations; and
(ii) the polynomial scalling is sub-optimal, a phenomenon known as critical slowing down.
We show that critical slowing down appears because of the limited possibilities to issue a local proposal on the Monte Carlo procedure in chaotic systems.
These results remain valid in other methods and show how generic properties of chaotic systems limit the efficiency of Monte Carlo simulations.
\end{abstract}

\pacs{05.10.Ln, 05.45.Pq}
\maketitle

\section{Introduction}

Caotic systems are characterized by quantities describing (statistical) properties of ensemble of trajectories, e.g. the Lyapunov exponent $\lambda$~\cite{OttBook} and the escape rate~$\kappa$ in open systems~\cite{LaiTamasBook}.
In any reasonably complicated system these quantities are estimated numerically from the integration of an ensemble of initial conditions $x$ for a finite time $N$.
Different $x$ lead to different estimations, e.g., they have different finite-time Lyapunov exponents $\lambda_N(x)$.
Indeed, the distribution of $\lambda_N(x)$ over randomly chosen initial conditions (or computed over the invariant measure of the system) is a characterization of the system and has been used to characterize dynamical trapping~\cite{Grassberger1985a, Zaslavsky2002a, Szezech2005a, Artuso2009a, Manchein2012}, to test hyperbolicity of the system~\cite{Vallejo2013}, or to identify small KAM islands~\cite{Tomsovic2007}.
In all these applications, the difficulty is to reliably estimate the tails of the distribution, which typically decay exponentially with $\lambda$ and $N$~\cite{BeckBook}.
Rare trajectories play a similar role in all (difficult) simulations of chaotic dynamical systems (e.g., in the characterization of open chaotic systems rare long-lived trajectories are essential to estimate $\kappa$ and $\lambda$).

\begin{figure}[!ht]
\includegraphics[width=\linewidth]{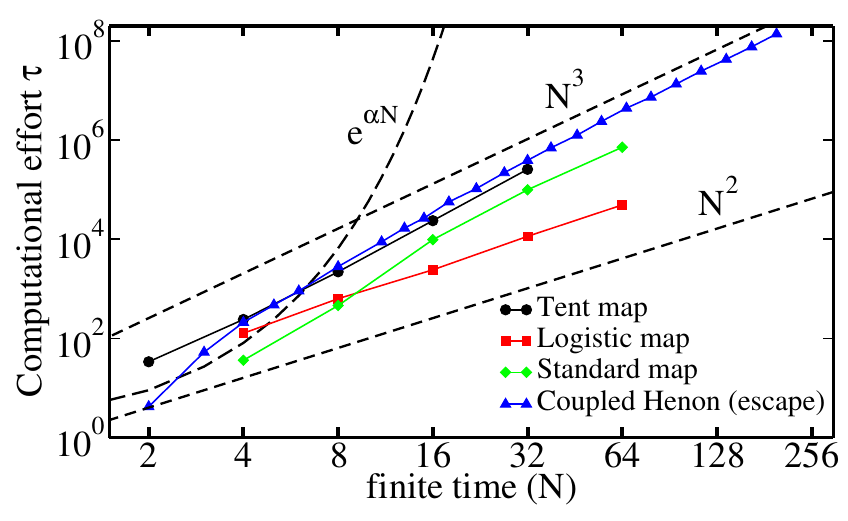}
\caption{
Computational effort of importance sampling Monte Carlo methods in chaotic systems.
Each curve represents the average round-trip time $\tau$ (a measure of the computational effort, see Sec.~\ref{sec:efficiency}) of a representative simulation (flat-histogram) on the distribution of the (largest) finite-time lyapunov exponent.
While uniform sampling scales as $\exp(\alpha N)$, importance sampling scales as $N^\gamma$ with $\gamma \geq 2$.
Tent map: Eq.~(\ref{eq:tent}) with $a=3$; Logistic map: $x_{n+1} = 4 x_n (1 - x_n)$; Standard map~\cite{OttBook}: $K=8$ and we used $\lambda_N(x_0, v_0)$ in Eq~(\ref{eq:lyapunov_nd}) where $v_0$ was drawn isotropically on every proposal;
Coupled Henon map: a 4 dimensional open system retrieved from Fig. 5 of Ref.~\cite{Leitao2013}, the simulation is on the escape time distribution.
The Wang-Landau algorithm~\cite{Wang2001} was used to estimate the distribution prior to perform the flat-histogram and the distribution agrees with the analytical one when available~\cite{Prasad1999, BeckBook}.
}
\label{fig1}
\end{figure}

Several methods have been proposed to find~\cite{Nusse1989, Sweet2001, Tailleur2007, Kitajima2011} and sample\cite{Kitajima2011, Philipp2010, Laffargue2013a, Leitao2013} rare trajectories in chaotic systems.
In particular, Lyapunov Weighted Dynamics~\cite{Tailleur2007, Laffargue2013a} and Lyapunov weighted path sampling~\cite{Philipp2010} use Monte Carlo importance sampling techniques to estimate the distribution of finite-time Lyapunov exponents.
A crucial ingredient in all these methods, and in Monte Carlo methods more generally, is the {\it locality} in the proposal: once a rare trajectory is found, it is essential to be able to propose another similarly rare trajectory.
Locality allows for a step-wise approximation of extremely rare trajectories.
Differently from other Monte Carlo applications in Physics, local proposals in the phase space of chaotic dynamical systems are not easy to obtain due to the exponential sensitivity of trajectories and fractal structures in the phase space~\cite{Leitao2013}.
While the success of the methods mentioned above indicates that it is possible to achieve local proposals, the implications of the limited locality to the computational efficiency of the Monte Carlo method has not been systematically explored yet.

In Fig.~\ref{fig1} we show how the computational effort of a representative Monte Carlo method (a flat-histogram simulation) scales with system size (e.g., in the computation of the distribution of Lyapunov exponents $\lambda_N(x)$ the role of system size is played by the finite-time $N$).
For different chaotic systems and problems, we obtain a polynomial scaling $\sim
N^\gamma$.
This is dramatically better than the exponential scaling $\sim e^{\alpha N}$ obtained using uniform sampling but is systematically worst than the theoretical optimal scaling $\sim N^2$ (a phenomenon known as critical slowing down~\cite{Trebst2006}).

The goal of this paper is to investigate the efficiency of importance sampling Monte Carlo methods and the origin of critical slowing down in chaotic dynamical systems.
After a general introduction to notions of chaotic systems and Monte Carlo methods (in Sec.~\ref{sec:importance-sampling}), we focus (in Sec.~\ref{sec:sequences}) on a flat-histogram Monte Carlo simulation of the distribution of $\lambda_N(x)$ in a simple dynamical system.
We obtain analytical estimations of the efficiency of this simulation (in Sec.~\ref{sec:subdiffusion}) which show that critical slowing down originates from the
interplay between chaotic properties and limitations imposed on the proposals.
We finish (in Sec.~\ref{sec:conclusions}) with a discussion of the implications of our results on Monte Carlo simulations in chaotic systems more generally.

\section{Importance Sampling in Chaotic systems \label{sec:importance-sampling}}

\subsection{Chaotic systems}

Let $x$ be a state in the phase-space $\Omega$ of a discrete time chaotic dynamical system defined by $x_{n+1} = F(x_n)$.
The finite-time Lyapunov exponent of a $N$-time trajectory starting at $x_0$ and for a unitary vector $v_0$,
\begin{equation}
\lambda_N(x_0, v_0) \equiv \frac{1}{N}\log | DF^t(x_0)\cdot v_0| \ \ ,
\label{eq:lyapunov_nd}
\end{equation}
where $DF^N$ is the derivative of $F$ composed $N$ times. For 1 dimension, Eq.~\ref{eq:lyapunov_nd} can be simplified to
\begin{equation}
\lambda_N(x_0) \equiv \frac{1}{N}\sum_{n=1}^N \log | F'(x_n)| \ \ ,
\label{eq:lyapunov}
\end{equation}
which measures the finite-time exponential divergence of two initial conditions starting close to $x_0$.
The distribution of $\lambda_N$,
\begin{equation}
\rho_N(\lambda) \equiv \frac{1}{V(\Omega)}\int_\Omega \delta(\lambda - \lambda_N(x))dx \ \ ,
\label{eq:distribution}
\end{equation}
measures the relative number of trajectories with a $\lambda_N$ between $\lambda$ and $\lambda + d\lambda$, where $V(\Omega)$ is the volume of $\Omega$.
We are mainly interested to sample states with different chaoticities to compute $\rho_N(\lambda)$ for a finite but high $N$.

\subsection{Importance sampling}

To sample states with different chaoticities, one needs to sample $\lambda$s from possible values in $[\lambda_{min}, \lambda_{max}]$, where $\lambda_{min}/\lambda_{max}$ are either the extreme values of $\lambda$ for a particular system, or a pre-selected region of $\lambda$s where we want to focus on (e.g. strongly chaotic trajectories only).
The fraction of samples in a given bin $b_\lambda=[\lambda, \lambda + \delta\lambda]\subset [\lambda_{min}, \lambda_{max}]$ is an estimate of $\rho_N(\lambda\in b_\lambda)$.
The variance $\sigma^2_{b_\lambda}$ of this estimate scales with the number of samples $M_{b_\lambda}$ on that bin~\footnote{We use '$\sim$' to mean 'scales as' and '$\approx$' to mean 'approximated to'.} as:

\begin{equation}
\sigma^2_b \sim \frac{1}{M_{b_\lambda}}\ \ .
\label{eq:variance}
\end{equation}
A uniform sampling consists in generating an ensemble of independent and uniformly-distributed initial conditions $x_i$, $i=1,...,M$ in the phase-space.
For each $x_i$, we compute $\lambda_N(x_i)$ from Eq.~(\ref{eq:lyapunov}) and estimate $\rho_N(\lambda\in b_\lambda) = M_{b_\lambda}/M$.
Because in a uniform sampling the probability to sample $\lambda \in b_\lambda$ is proportional to $\rho_N(\lambda)$, on average
\begin{equation}
M_{b_\lambda} = \rho(\lambda) M
\label{eq:samples_uniform}
\end{equation}
and Eq.~(\ref{eq:variance}) yields
\begin{equation}
\sigma^2_{b_\lambda} \sim \frac{1}{\rho_N(\lambda) M}\ \ .
\label{eq:uniform}
\end{equation}
In fully chaotic systems, the tails of $\rho_N(\lambda)$ typically decrease exponentially as a function of $\lambda$ and of N~\cite{BeckBook, Prasad1999} and therefore $\rho_N(\lambda)$ has exponentially high variance.
To compensate the high variance, a uniform sampling requires exponentially high number of samples $M$ with increasing N and $\lambda$.
I.e. for a fixed accuracy $\sigma^2_{b_\lambda}$ of $\rho_N(\lambda)$, the computational effort of an uniform sampling increases exponentially with N.

Importance sampling techniques aim to reduce $\sigma^2_{b_\lambda}$ by sampling from a non-uniform distribution $P(x)$ such that rare $\lambda$s (in the tails of $\rho_N(\lambda)$) are more often sampled, i.e., they generate samples in such a way that $M_{b_\lambda}$ is no longer given by Eq.~(\ref{eq:samples_uniform}).
One example is the canonic ensemble~\cite{NewmanBarkemaBook}, proposed to sample chaotic systems in Refs.~\cite{Tailleur2007, Philipp2010},
\begin{equation}
P(x) = P_{\beta}(x) \equiv \frac{1}{Z_\beta}e^{-\beta \lambda_N(x)}\ \ ,
\label{eq:canonic}
\end{equation}
where $Z_\beta \equiv \int_\Omega dx e^{-\beta \lambda_N(x)}$ and $\beta$ is chosen to increase $M_{b_\lambda}$ for specific $\lambda$s.
Specifically, we have that
\begin{equation}
M_{b_\lambda} = M P_{\beta}(\lambda) \rho_N(\lambda) \sim M e^{-\beta \lambda + S_N(\lambda)}
\label{eq:sampling_canonic}
\end{equation}
where $S_N(\lambda)\equiv \log(\rho_N(\lambda))$.
By choosing specific values of $\beta$, more importance is given to specific $\lambda$s centered in the maximum of $M_{b_\lambda}$ in Eq.~(\ref{eq:sampling_canonic}), obtained implicitly from the solution of $dS_N(\lambda)/d\lambda=\beta$.
A re-weighting technique~\cite{Ferrenberg1988} can be used to compute $\rho_N(t)$ from canonic ensemble simulations with different $\beta$s.

Here we focus on the flat-histogram ensemble~\cite{Berg1991}, that has been used in chaotic systems in Refs.~\cite{Kitajima2011, Leitao2013},
\begin{equation}
P(x) = P_f(x) \equiv \frac{Z}{\rho_N(\lambda(x))}\ \ ,
\label{eq:flat-histogram}
\end{equation}
where $Z$ is a normalization constant (if $\rho_N(t)$ is unknown, the Wang-Landau algorithm was used~\cite{Wang2001}).
This choice ensures $M_{b_\lambda}$ is independent of $\lambda$, i.e. 
\begin{equation}
M_{b_\lambda} = M / \#\text{bins}
\label{eq:sampling_multicanonic}
\end{equation}
and thus $\sigma^2_{b_\lambda} $ in Eq.~(\ref{eq:variance}) is independent of $\rho_N(\lambda)$.

\subsection{Markov Chain}

To draw initial conditions $x_i$ from the distribution $P(x)$, a Metropolis-Hastings algorithm is used.
It is a Markov chain with a transition probability from a state $x$ to a state $x'$ given by $P(x\rightarrow x') = g(x\rightarrow x') A(x \rightarrow x')$, where $g(x\rightarrow x')$ is the conditional probability to propose $x'$ given $x$, and $A(x \rightarrow x')$ is the conditional probability to accept $x'$ given $x$~\cite{NewmanBarkemaBook}.
Any initial distribution asymptotically converges to $P(x)$ if the chain is ergodic and satisfies detailed balance

\begin{equation}
P(x\rightarrow x')P(x) = P(x'\rightarrow x)P(x')\ \ ,
\label{eq:balance}
\end{equation}
which is fulfilled using the Metropolis choice~\cite{NewmanBarkemaBook}:
\begin{equation}
A(x \rightarrow x') = \min \left\{1, \frac{P(x')}{P(x)} \frac{g(x'\rightarrow x)}{g(x \rightarrow x')}  \right\}\ \ .
\label{eq:acceptance}
\end{equation}

The proposal distribution $g(x\rightarrow x')$ is a free parameter of any Metropolis-Hastings algorithm and has to be adjusted according to the problem.
Ideally, the proposal should be chosen to maximize the mobility of the simulation on the phase-space.
On one hand, it has to be able to propose $x'$ such that $\lambda_N(x')$ is far enough from $\lambda_N(x)$ that the simulation visits all $\lambda\in[\lambda_{min}, \lambda_{max}]$.
On the other hand, $\lambda_N(x')$ must be close enough from $\lambda_N(x)$ for the ratio $P(x')/P(x) = P(\lambda_N(x'))/P(\lambda_N(x))$ in Eq.~(\ref{eq:acceptance}) be high enough that $x'$ is accepted and the simulation is able to move.
These conditions are satisfied issuing local proposals~\cite{Berg1991, Leitao2013}, i.e. proposals that change $\lambda$ only by a small amount:
\begin{equation}
N\delta_\lambda \equiv N|\lambda_N(x) - \lambda_N(x')| \approx 1\ \ .
\label{eq:locality}
\end{equation}
In the literature two options have been suggested to obtain local steps in simulations of chaotic systems:
\begin{itemize}
\item  \em{}shift\em{}~\cite{Dellago2002} proposes a state $x'$ by a forward or backward iteration of $x$ (i.e. $x' = F(x)$ or $x' = F^{-1}(x)$)
\item  \em{}precision shooting\em{}~\cite{Grunwald2008} proposes $x'$ in a neighborhood $\delta(x)$ of $x$ ($x' = x + \delta(x)$). The critical step is that $\delta(x)$ has to decay exponentially as~\cite{Grunwald2008, Leitao2013}
\end{itemize}
\begin{equation}
\delta(x) = \delta_0 e^{-\lambda_N(x) N}\ \ .
\label{eq:local_sigma}
\end{equation}
The shift fulfills Eq.~(\ref{eq:locality}) because it only changes one term on the sum of Eq.~(\ref{eq:lyapunov}) and thus $|\lambda_N(x) - \lambda_N(x')| \approx 1/N$.
The precision shooting fulfills Eq.~(\ref{eq:locality}) because after $N$ map iterations the exponential sensitivity of initial conditions leads to $|F^t(x)-F^t(x')|$ be roughly $\delta_0$ and thus, by choosing $\delta_0$ in Eq.~\ref{eq:local_sigma}, we can make the trajectories to be as close as we want and thus make $N|\lambda_N(x) - \lambda_N(x')|$ to be as close as we want from 1.

As in Ref.~\cite{Grunwald2008}, we consider here a mixed of the two proposals.
If we were to use just precision shooting, the step-size in the phase-space would be exponentially small as $N\rightarrow \infty$ and the simulation would be stuck in $\lambda$.
On the other hand, if we were to use just the shift, we would be moving forward and backward along a particular trajectory; while it would be a valid proposal, it would never be better than just iterating the trajectory forward (and performing a time-average of it).
In other words, the shift moves to different regions of the phase-space and the precision shooting randomizes the trajectory and both are required for an efficient simulation.

In summary, the algorithm consists in starting a random initial condition $x$ and:
\begin{enumerate}
       \item Iterate $x$ for $N$ times and compute $\lambda$ using Eq.~(\ref{eq:lyapunov_nd});
       \item Propose a state $x'$ using precision shooting with 1/2 probability, 1/4 using forward shift, and 1/4 using backward shift: $g(x\rightarrow x') = \frac{1}{2\delta(x)}\exp(-|x-x'|/\delta(x))$ $+ 1/4\delta(x - F(x))$ $+ 1/(4I)\sum_i^I \delta(x - F_i^{-1}(x))$ where the sum is over all pre-images $I$ (important when the map is not invertible);
       \item Accept or reject $x'$ according to the probability $A(x \rightarrow x')$, Eq.~(\ref{eq:acceptance}), where $g(x\rightarrow x')$ is the proposal made on the previous step and $P(x) = P_f(x)$ of Eq.~(\ref{eq:flat-histogram}).
       \item go to 2.
\end{enumerate}

\subsection{Efficiency\label{sec:efficiency}}

To compare the efficiency of different methods fairly, we need to take into account that samples obtained from Markov chains are typically correlated.
Thus, we measure how the number of chain iterations required to obtain an independent sample on any bin $\tau_s$ scales with $N$.
In simulations with local proposals, such as ours or single spin simulations in lattices, $\tau_s$ is difficult to estimate and typically the round-trip time $\tau$ is used as a proxy of $\tau_s$ to measure the computational effort~\cite{Dayal2004, Costa2007, Leitao2013}.
The round-trip time is defined as average number of chain iterations required to bring any state $x$ with $\lambda_N(x) = \lambda_{max}$ to any state $x$ with $\lambda_N(x) = \lambda_{min}$ and return back (i.e. the path $\lambda_{max}\rightarrow \lambda_{min}\rightarrow \lambda_{max}$).

In uniform sampling, the round-trip time can be estimated by the number of samples $M$ required to obtain one state $x$ with $\lambda_N(x) = \lambda_{min}$.
This is proportional to $\rho_N(\lambda_{min})$ and we thus recover the same scalling as we have obtained using Eq.~(\ref{eq:uniform}).
To understand how the round-trip time scales in a flat-histogram simulation, it is constructive to look at the Markov chain projected on $\lambda$ as a random walk on the real line $[\lambda_{min}, \lambda_{max}]$.
A simulation starting in $x$ slowly walks in $\lambda$ (by transiting to $x$ with other $\lambda$s) because the step-size fulfills Eq.~(\ref{eq:locality}). In the best case, the average displacement in $N\lambda$ after t iterations is $\sigma_\lambda=\sqrt{t}$.
In this case, to displace the full interval $N(\lambda_{max} - \lambda_{min})\sim N$~\footnote{We use that $\lambda$, and in particular $\lambda_{max}$ and $\lambda_{min}$, are intensive variables~\cite{BeckBook}.}, we require a number of steps $\tau$ such that $\sigma_\lambda(\tau) \sim N$, or
\begin{equation}
\tau \sim N^2\ \ .
\label{eq:diffusive_roundtrip}
\end{equation}

In Fig.~\ref{fig1} we plot different numerically computed round-trip time $\tau$ as a function of the finite-time $N$ for different systems and see that the scalling of Eq.~(\ref{eq:diffusive_roundtrip}) is not observed in all cases.
This phenomena is known in the literature of Monte Carlo in spin systems as critical slowing down~\cite{Trebst2006}.
It occurs even in simple spin systems such as the 2D Ising model and attempts have been made to explain it~\cite{Dayal2004, Costa2007}.
In order to understand the origins of critical slowing down in chaotic systems, we use a simple system where analytical calculations of $\tau_s$ and $\tau$ can be performed.

\section{Simulations on the Tent map\label{sec:sequences}}

\subsection{Tent map}

\begin{figure}[!ht]
\includegraphics[width=\linewidth]{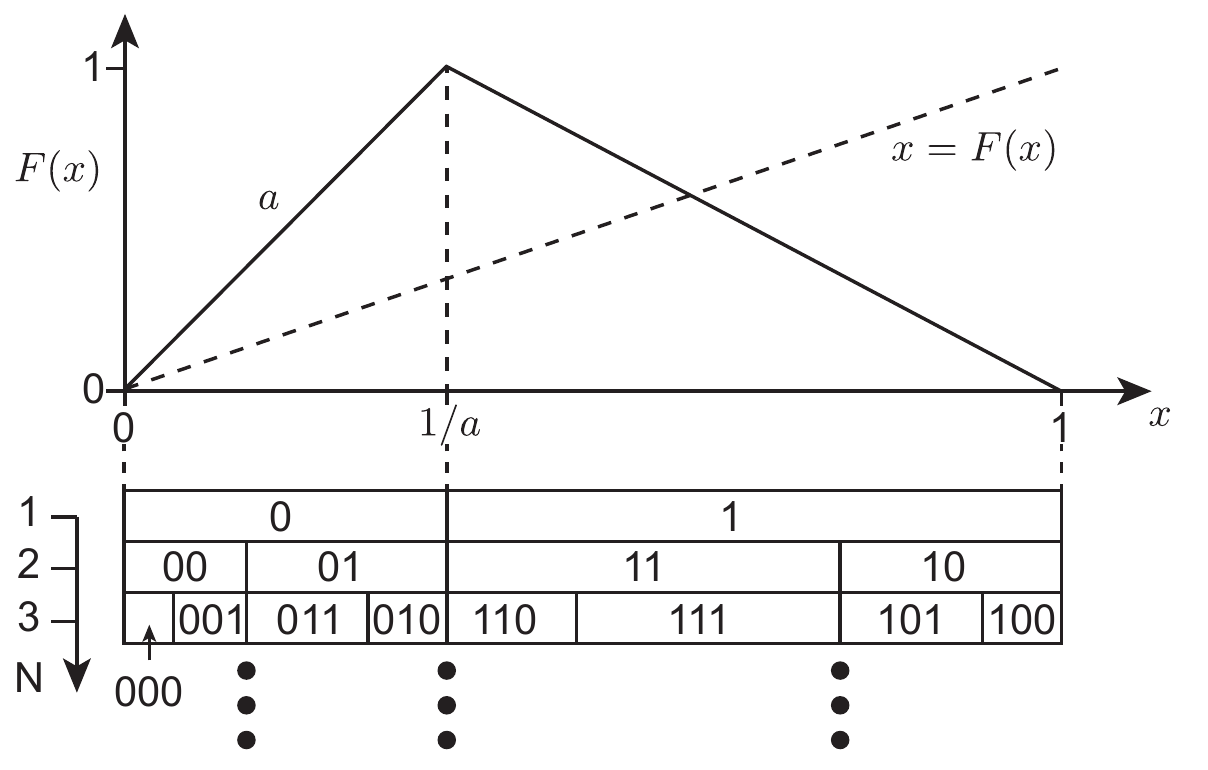}
\caption{
The two-scale tent map, Eq.~(\ref{eq:tent}), and its finite-time $N$ symbolic sequences $s=[s_1,...,s_N]$.
}
\label{fig2}
\end{figure}

We consider the paradigmatic one dimensional two-scale tent map, defined in the interval $[0,1]$ for $a\geq 2$ by the equation $x_{n+1} = F(x_n)$ with

\begin{equation}
F\left(x\right)=\begin{cases}
a x & \text{if}\,\, 0\leq x\leq \frac{1}{a}\\
\frac{a}{a-1}\left(1-x\right) & \text{if}\,\, \frac{1}{a}<x\leq 1\\
\end{cases}
\label{eq:tent}
\end{equation}
as shown in Fig.~\ref{fig2}.
We can construct a symbolic representation of the system by considering its natural Markov partition in two intervals $I_0 \equiv[0,1/a]$ and $I_1\equiv[1/a,1]$.
Any trajectory starting at $x_0$ can be represented by a symbolic sequence $s(x_0)=[s_1 s_2 s_3 ... s_N]$, where $s_i=0 \text{ if } x_i\in I_0$ and $s_i=1 \text{ if } x_i\in I_1$.
There are $2^N$ different symbolic sequences and $N+1$ different possible values of the finite-time Lyapunov exponent (Eq.~\ref{eq:lyapunov}), which in this case depends only on the number $k$ of 0's in the sequence $s$,
\begin{equation}
\lambda_{N,k}=\frac{1}{N}\left(k\log a+\left(N-k\right)\log\frac{a}{a-1}\right)\,\, k=0,...,N\, \, .
\label{eq:lambda_tent}
\end{equation}
The relative number of sequences with a given number $k$ of 0's is
\begin{eqnarray}
\rho_{s}\left(\lambda_{N,k} \right) & = & \frac{1}{2^{N}}\binom{N}{k}\ \ ,
\label{eq:rho_sequences}
\end{eqnarray}
where $\binom{N}{k}\equiv N!/((N-k)!k!)$.
The number of (phase-space) states with the same symbolic sequence $s$ is given by the measure $\mu_{N,k}$ of the phase-space interval of $s$.
In this system, the measure is uniform and therefore $\mu_{N,k}$ equals the length of each interval, which is given by $(1/a)^k((a-1)/a)^{N-k}$.
Thus, the number of states with a given $\lambda_{N,k}$ is
\begin{equation}
\rho\left(\lambda_{N,k} \right) = \rho_{s}\left(\lambda_{N,k} \right) \mu_{N,k} = \frac{1}{2^{N}}\binom{N}{k}\frac{1}{a^{k}}\frac{a-1}{a}^{N-k}\, \, .
\label{eq:rho_tent}
\end{equation}

\subsection{Monte Carlo on the symbolic sequences}

Analytical calculations of how the computational effort of a Monte Carlo simulation scales with the system size are typically impossible because of the complexity of the underlying Markov chain~\cite{RobertCasellaBook}.
In applications to dynamical systems, this problem is even more difficult because the Markov chain is defined on a continuous phase-space.
The advantage of the tent map is that we can map the Monte Carlo simulation in its phase-space to a Monte Carlo simulation in the (discrete) space of symbolic sequences and analytically treat this simpler simulation.
Specifically, we introduce a new Monte Carlo process defined on the set of all possible symbolic sequences of the tent map ($2^N$) that moves through a standard discrete-space Metropolis-Hastings process: from $s$ we propose a sequence $s'$ with a given proposal distribution and we accept or reject it according to an acceptance distribution.
We thus have two distinct simulations:
\begin{enumerate}
\item on the phase-space, using $x\in \Omega=[0,1]$ with the procedure outlined in the previous section;
\item on the symbolic sequences, with $\Omega$ the set of all binary sequences.
\end{enumerate}
The crucial step is to construct the simulation 2. in such a way that it is equivalent to simulation 1.

\begin{figure}[!ht]
\includegraphics[width=\linewidth]{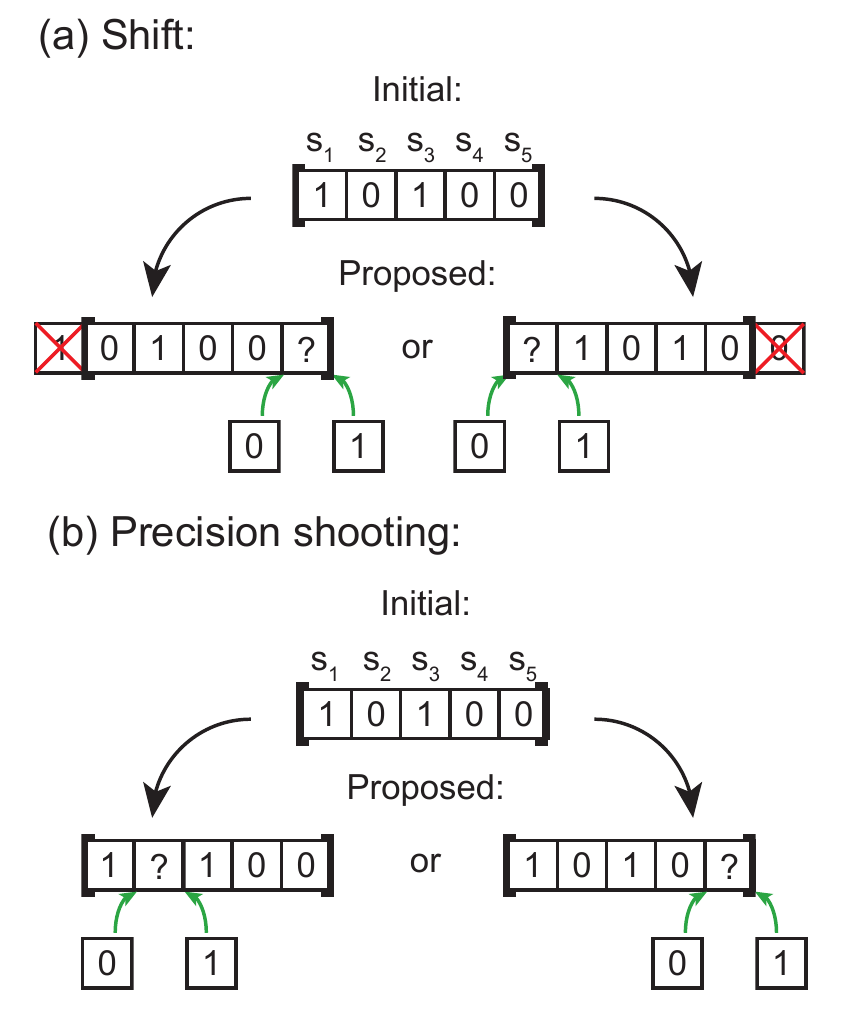}
\caption{
Local proposals in the symbolic sequences of the tent map.
(a) The {\em shift} proposes to shift the sequence, where the last (first) symbol is dropped and a new symbol is added in the beginning (end).
(b) The {\em precision shooting} proposes a change in the symbol after the first "1" when counting from the right (left panel), or in the last symbol (right panel).
}
\label{fig3}
\end{figure}

We first map the proposals in simulation 1. to proposals in simulation 2.:
\begin{itemize}
\item the {\em shift} corresponds to have the whole sequence $s(x)$ shifted by one symbol, where the last symbol is dropped and a new symbol $s$ is added in the beginning (or the opposite to the backward shift), see Fig.~\ref{fig3}(a). The new symbol $s_i$ (0 or 1) appears with probability $\mu_{s_i}$ to correspond to the respective measure of the phase-space \footnote{On the phase-space, a forward shift followed by a backward shift sends the state $x$ exactly to the same state it was two steps before because the system is deterministic.
Our approximation is that the randomization due to precision shooting in simulation 1. leads to to the new symbol $s_i$ to appear with probability $\mu_{s_i}$ as if the initial condition $x$ with symbolic sequence $s(x)$ would be in any position $y$ such that $s(y) = s(x)$.}.
\item {\em precision shooting} corresponds to propose either to the same symbolic sequence $s$ or a neighbor sequence {\em on the phase-space}. E.g. from $s=[011]$ in Fig.~\ref{fig2}, it proposes $s'=[011]$, $s'=[001]$, or $s'=[010]$. For the tent map, this can be written as a simple rule, see Fig.~\ref{fig3}(b).
\end{itemize}

The acceptance (Eq.~(\ref{eq:acceptance}) with $x$ replaced by $s$) is mapped by taking into account that the proposal is symmetric (i.e. $g(s\rightarrow s') = g(s'\rightarrow s)$) and that $\lambda$ of $s$ is computed using Eq.~(\ref{eq:lambda_tent}) by counting the number of 0's in $s$.

\begin{figure}[!ht]
\includegraphics[width=\linewidth]{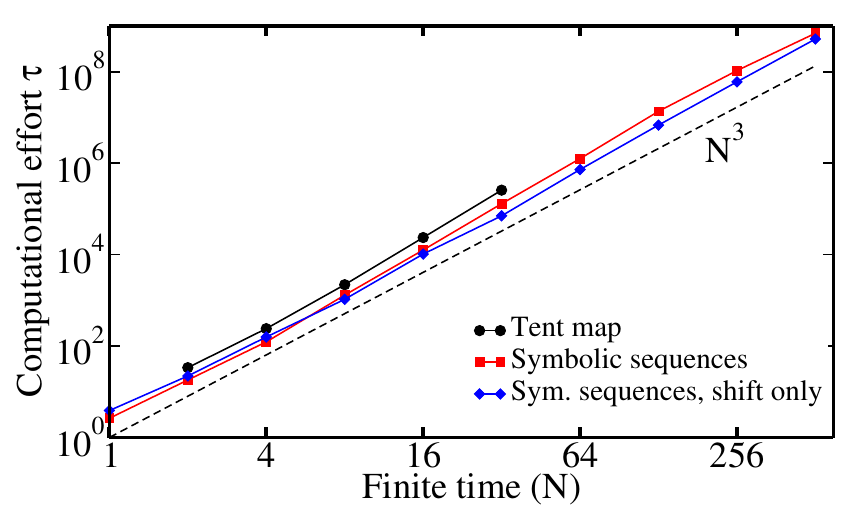}
\caption{
Equivalence between a Monte Carlo on the tent map and on its symbolic sequences.
Scaling of the average round-trip as function of the finite-time $N$ of a flat-histogram simulation on the tent map (black circles) and on its symbolic sequences (red rectangles).
In blue diamonds is the same flat-histogram simulation in symbolic sequences, but only using shift proposals.
Dashed line represent the scaling $N^3$.
We have used $a=3$ in Eq.~(\ref{eq:tent}) and, for the simulation on the phase-space, we have used the procedure outlined in the previous section with the exact distribution Eq.~(\ref{eq:rho_tent}) in Eq.~(\ref{eq:flat-histogram}), $\delta_0=0.1$ in Eq.~(\ref{eq:local_sigma}), and the average round-trip was computed over 100 round-trips.
In simulations on the symbolic sequences, we have used 50\% probability for the same sequence, and 25\% for each neighbor interval on the precision shooting, but we observe no qualitative difference with other values.
}
\label{fig4}
\end{figure}

In Fig.~\ref{fig4} we compare simulations 1. and 2. and observe no quantitative difference in the scalling of the round-trip time of both simulations, indicating that they are indeed equivalent.
We observe a critical slowing down with scalling $\tau \sim N^3$ over more than two decades, which we aim to explain in the next section.
Furthermore, by comparing the simulation 2. with and without precision shooting, we see that the effect of precision shooting is to move $\tau$ vertically (making simulations less efficient) and seems to have no effect on the scaling; we thus neglect it on the next section.
In simulation 1., precision shooting is essential and we can neglect in simulation 2. because we have used it in simulation 1.~[31].

\section{Explanation of sub-optimal scaling\label{sec:subdiffusion}}

To explain sub-optimal scalling, we need to consider an ensemble of independent simulations and compute how the average round-trip time $\tau$ scales with increasing $N$.
In Sec. \ref{sec:importance-sampling} we presented an expression for the round-trip time (Eq.~(\ref{eq:diffusive_roundtrip})) based on the assumption that the ensemble diffuses in $\lambda\in[\lambda_{min},\lambda_{max}]$ with a variance $\sigma_\lambda \sim \sqrt{t}$. Critical slowing down shown in Fig.~\ref{fig4} shows that this assumption is wrong for this case. Here drop this assumption and compute explicitly how $\sigma_\lambda$ evolves with $t$.
A simulation on the symbolic sequences with only shifts is equivalent to a window of size $N$ moving on a tape of 0's and 1's that, at each step, moves to the left or to the right and randomizes the symbol that enters the window (see Fig.~\ref{fig3}a, where the boundaries of the window are the right brackets in bold)~\footnote{The assumption that the window moves randomly to the left and to the right with equal probabilities is an approximation because it neglects the acceptance probability.
This approximation is justified because the acceptance in simulation 2. only depends on the ration $P(s')/P(s)$, which only depends on symbol $s_0$ or $s_N$ that changes in $s$. By symmetry, it is equally likely to be $s_0$ or $s_N$ to change and thus the process is symmetric in respect to left or right.}.
$\lambda$ is proportional to the sum of 1's in the window (from Eq.~(\ref{eq:lambda_tent})) and, in particular, $\lambda_{min}$ occurs in the sequence [00...0] and $\lambda_{max}$ in the sequence [11...1].
Our problem is to compute how the round-trip time $\tau$ --- the number of window moves required to complete the path $[00...0]\rightarrow [11...1] \rightarrow [00...0]$ --- scales with $N$.

We first note that the window performs a simple random walk and thus the number of symbols $\Delta(t)$ that changed after $t$ Monte Carlo steps scale as
\begin{equation}
\Delta(t) \sim \sqrt{t}\ \ .
 \label{eq:variance_sequence}
\end{equation}
Since, on average, at time $t$ only $\Delta(t)$ of the $N$ symbols changed, the variance $\sigma^2_\lambda(t)$ can only depend on the symbols that changed. Since $\lambda$ is proportional to the sum of symbols by Eq.~(\ref{eq:lambda_tent}), its variance is proportional to the number of symbols that changed $\Delta(t)$, and we obtain:

\begin{equation}
\sigma_\lambda^2(t) \sim \Delta(t) \sim \sqrt{t}\ \ ,
 \label{eq:sigma_lambda}
\end{equation}
that confirms the existence of a subdiffusion in $\lambda$.
The average time to obtain an independent sequence $s$ is the time $\tau_s$ such that all symbols have changed, or $\Delta(\tau_s) \approx N$. From Eq.~(\ref{eq:sigma_lambda}) we obtain
\begin{equation}
\tau_s \sim N^2\ \ .
 \label{eq:decorrelation_time}
\end{equation}

To appreciate the relevance of this result we have to compare it to how would $\tau_s$ scale if we were able to randomly change any symbol of the symbolic sequence $s$ at each time.
Because $s$ has $N$ symbols, we would require $t = N$ steps and thus $\tau_s \sim N$, which would not have any critical slowing down.
Eq.~(\ref{eq:decorrelation_time}) thus indicates that the proposal derived to correspond to the proposals in phase-space dramatically limit our allowed moves and changes the scaling of $\tau_s$.
We can summarize the above results in the following picture: on a time-scale up to $\tau_s$ given by Eq.~(\ref{eq:decorrelation_time}), the random walk in $\lambda$ subdiffuses according to Eq.~(\ref{eq:sigma_lambda}); on a larger time-scale, the random walk diffuses normally as it draws independent sequences.
\begin{figure}[!ht]
\includegraphics[width=\linewidth]{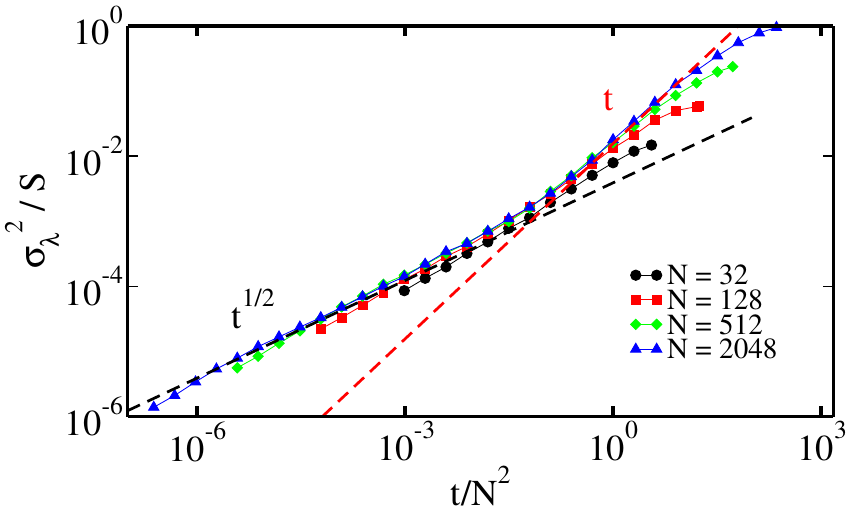}
\caption{
Critical slowing down in Monte Carlo simulations on the symbolic sequences of the tent map.
The variance $\sigma_\lambda^2(t)$ was estimated as $1/M\sum_{i=1}^M (\lambda_i(t) - \overline{\lambda}(t))^2$ where $\overline{\lambda}(t) =\sum_{i=1}^M \lambda_i(t)$ and M is number of independent simulations.
We divided it by the asymptotic variance for a flat-histogram divided by $N$, $S=[(N+1)/12]/N$; the x-axis represents the number of Monte Carlo steps divided by $N^2$.
Different curves represent averages over $M=1000$ independent flat-histogram simulations starting from states $s$ with a fixed $\lambda_0==\lambda_{N,N/2}$ (see Eq.~(\ref{eq:lambda_tent})) for different finite-times from $N=32$ up to $N=2048$.
Two scalings, $\sqrt{t}$ and $t$ are shown in dashed.
Changing $\lambda_0$ does not change the shape, only shifts all curves.
}
\label{fig5}
\end{figure}
Results shown in Fig.~(\ref{fig5}) confirm this picture as we observe: a) a transition from $\sigma^2_\lambda \sim t^\frac{1}{2}$ to $\sigma_\lambda^2 \sim t$; b) the transition occurs at a transition time independent of $N$ when time is rescaled by $1/N^2$, as predicted by Eq.~(\ref{eq:decorrelation_time}).

We now compute the round-trip time $\tau$. Consider the Markov process obtained as $\tau_s$ iterations of the original process. The original time $t$ relates to the new time $t'$ by $t'=t/\tau_s$.
The new process generates independent symbolic sequences and makes steps in $\lambda$ of size $\sigma_\lambda(\tau_s) \sim \sqrt{N}$.
This implies that the process now diffuses normally in $\lambda$ with a variance $\sigma^2_{\tau_s}(t')$ given by
\begin{equation}
\sigma^2_{\tau_s}(t')=2D t' \sim \sigma_\lambda^2(\tau_s) t' \sim t/N
 \label{eq:sigma_tun}
\end{equation}
where we used $t'=t/\tau_s$, Eqs.~(\ref{eq:sigma_lambda})-(\ref{eq:decorrelation_time}), and that the diffusion coefficient of a random walk with step-size normally distributed is proportional to $\sigma^2_\lambda(\tau_s)$.
As argued before, performing a round-trip corresponds to $\sigma_{\tau_s}(\tau) \approx N$. Using Eq.~(\ref{eq:sigma_tun}), can obtain
\begin{equation}
\tau \sim N^3\ \ ,
\end{equation}
in agreement with the scaling observed in Fig.~\ref{fig4}.

\section{Conclusions\label{sec:conclusions}}

In this paper we described how importance sampling Monte Carlo methods can improve simulations in chaotic systems, in particular to the problem of the computation of the distribution of finite-time Lyapunov exponent.
We numerically computed the efficiency of a flat-histogram simulation in different systems and verified that it outperforms uniform sampling: the scalling changes from exponential to polynomial.
However, the exponent of the polynom is not $2$ as we would expect from a simple random walk on a real line, a phenomena known in the literature of spin systems as critical slowing down.
Using a simple system that presents critical slowing down, we analytically showed that in this system the Markov process decorrelates with $\tau_s \sim N^2$ due to a subdifussion on the finite-time Lyapunov exponent.
This allowed us to derive the scalling of the round-trip time as $\tau \sim N^3$, in excellent agreement with the simulations.
To our knowledge, this is the first time the scaling of round-trip time of a Monte Carlo simulation was analyticaly computed in a non-trivial system.

The importance of our results is not limited to flat-histogram simulations. The sub-optimal sampling we observe is a direct consequence of the limited options of local proposals that can be generated in chaotic systems and therefore should affect all importance sampling Monte Carlo simulations using these proposals. For instance, canonic simulations using these proposals-- which have been used before for estimating the distribution of finite-time Lyapunov exponents~\cite{Philipp2010, Laffargue2013a} -- in the tent map should follow the scaling $\tau_s \sim N^2$ derived in Eq.~(\ref{eq:decorrelation_time}). While extending the validity of the scalings derived in this paper to other methods is not straightforward, our results show the need of a more careful investigation of the efficiency of modern computational methods applied to dynamical systems.

The efficiency of the simulations is not only a property of the specific Monte Carlo method, it is a result of an interplay between the method (e.g., the proposals) and the phase space structures of the chaotic system (e.g., fractals). Indeed, we had already observed sub-optimal scaling of the efficiency in our previous work on open chaotic systems~\cite{Leitao2013}. The importance of this paper is to show one mechanism for such sub-optimal scaling. 
Even if different scalings should be expected for different problems and systems (see Fig. 1), the mechanisms reported here is expected to affect also more general classes of dynamical systems.

J.C.L. acknowledges funding from Funda\c{c}\~{a}o para a Ci\^{e}ncia e Tecnologia (Portugal),  grant SFRH/BD/90050/2012.

\end{document}